\documentclass[12pt,preprint]{aastex}

\singlespace

\usepackage{natbib}
\bibpunct{(}{)}{,}{a}{}{,}

\newcommand{\lya}{Ly$\alpha$ }

\def\ltsima{$\; \buildrel < \over \sim \;$}
\def\simlt{\lower.5ex\hbox{\ltsima}}
\def\gtsima{$\; \buildrel > \over \sim \;$}
\def\simgt{\lower.5ex\hbox{\gtsima}}

\slugcomment{}

\shorttitle{\lya blobs at $z=3.1$}
\shortauthors{Matsuda et al.}

\begin{document}

\title{A Subaru SEARCH FOR \lya BLOBS IN AND AROUND THE PROTO-CLUSTER REGION AT REDSHIFT $z=3.1$ \altaffilmark{1} }

\author{Yuichi Matsuda      \altaffilmark{2,3},
        Toru Yamada         \altaffilmark{2},
        Tomoki Hayashino    \altaffilmark{3},
        Hajime Tamura       \altaffilmark{3},
        Ryosuke Yamauchi    \altaffilmark{3},
        Masaru Ajiki        \altaffilmark{4},
        Shinobu S. Fujita   \altaffilmark{4}, 
        Takashi Murayama    \altaffilmark{4},
        Tohru Nagao         \altaffilmark{4},
        Kouji Ohta          \altaffilmark{5},
        Sadanori Okamura    \altaffilmark{6,7},
        Masami Ouchi        \altaffilmark{6},
        Kazuhiro Shimasaku  \altaffilmark{6,7},
        Yasuhiro Shioya     \altaffilmark{4}, and 
        Yoshiaki Taniguchi  \altaffilmark{4}
}

\email{matsuda@awa.tohoku.ac.jp}

\altaffiltext{1}{Based on data collected at Subaru Telescope and in part obtained from data archive at Astronomical Data Analysis Center, which are operated by the National Astronomical Observatory of Japan.}
\altaffiltext{2}{National Astronomical Observatory of Japan, Mitaka, Tokyo 181-8588, Japan}
\altaffiltext{3}{Research Center for Neutrino Science, Graduate School of Science, Tohoku University, Aramaki, Aoba, Sendai 980-8578, Japan}
\altaffiltext{4}{Astronomical Institute, Graduate School of Science, Tohoku University, Aramaki, Aoba, Sendai 980-8578, Japan}
\altaffiltext{5}{Department of Astronomy, Kyoto University, Sakyo-ku, Kyoto 606-8502, Japan}
\altaffiltext{6}{Department of Astronomy, School of Science, University of Tokyo, Tokyo 113-0033, Japan}
\altaffiltext{7}{Research Center for the Early Universe, School of Science, University of Tokyo, Tokyo 113-0033, Japan}

\begin{abstract}

  We report the properties of the 35 robust candidates of \lya blobs (LABs), which are larger than 16 arcsec$^2$ in isophotal area and brighter than $0.7 \times 10^{-16}$ ergs s$^{-1}$ cm$^{-2}$, searched in and around the proto-cluster region at redshift $z=3.1$ discovered by Steidel et al. in the SSA22 field, based on wide-field ($31'\times23'$) and deep narrow-band ($NB497$; 4977/77) and broad-band ($B$,$V$, and $R$) images taken with the prime-focus camera on the Subaru telescope. The two previously known giant LABs are the most luminous and the largest ones in our survey volume of $1.3 \times 10^5$ Mpc$^3$. We revealed the internal structures of the two giant LABs and discovered some bubble-like features, which suggest that intensive starburst and galactic superwind phenomena occurred in these objects in the past. The rest 33 LABs have isophotal area of $\sim$16--78 arcsec$^2$ and flux of 0.7--7 $\times 10^{-16}$ ergs s$^{-1}$ cm$^{-2}$. These 35 LABs show a continuous distribution of isophotal area and emission line flux. The distributions of average surface brightness and morphology are widespread from relatively compact high surface brightness objects to very diffuse low surface brightness ones. The physical origins of these LABs may be (i) photo-ionization by massive stars, or active galactic nuclei, or  (ii) cooling radiation from gravitationally heated gas, or (iii) shock heating by starburst driven galactic superwind. One third of them are apparently not associated with ultra-violet continuum sources that are bright enough to produce \lya emission, assuming a Salpeter initial mass function. The 90$\%$ of these LABs are located inside the high surface density region of the 283 relatively compact and strong \lya emitters selected in our previous study. This suggests that these LABs may be the phenomena related to dense environment at high redshift. 

\end{abstract}


\section{INTRODUCTION}

 Recently, \lya imaging at high-redshift revealed the existence of very luminous and extended \lya nebulae, so-called \lya blobs (hereafter, LABs), which have the \lya luminosity of more than $10^{43}$ ergs s$^{-1}$ and the physical extent of about 100 kpc (e.g., Keel et al. 1999, hereafter K99; Steidel et al. 2000, hereafter S00; Francis et al. 2001, hereafter F01). These LABs are similar to the \lya halos often seen around powerful radio galaxies at high redshift, but they are not associated with luminous radio sources (K99; S00; F01). 

 The ionization or excitation mechanisms of these LABs are unclear. Of these, two (No.18 in K99 and 2142-4420 B1 in F01) are likely to have active galactic nuclei (AGNs), but other two (SSA22 blob1 and 2 in S00) show no evidence for AGNs although they are associated with Lyman break Galaxies (LBGs) at the same redshift. It was shown, however, that all these LABs do not have ultra-violet (UV) sources apparently bright enough to produce the extended photo-ionized \lya emission line nebulae. Until now, mainly three ideas are proposed to explain these LABs; (i) photo-ionization by UV sources obscured from our line of sight (e.g., Chapman et al. 2003), (ii) cooling radiation from gravitationally heated gas in collapsed halos (S00; Haiman, Spaans, \& Quataert 2000; Fardal et al. 2001), and  (iii) shock heating by starburst driven galactic superwind (Taniguchi \& Shioya 2000; Taniguchi, Shioya, \& Kakazu 2001; Ohyama et al. 2003). Scattering of Ly$\alpha$ photons by surrounding neutral gas may play an important role for all these cases. Since all these LABs mentioned above are claimed to lie in proto-cluster regions, it is possible that the phenomena are related to dense environment, where early galaxy formation occurred preferentially (Governato et al. 1998) and possibly rich in hydrogen gas (e.g., Adelberger et al. 2003). Recently, Palunas et al. (2004) also reported their detection of a few candidates of LABs associated with the overdensity region of the \lya emitters around 2142-4420 B1. On the other hand, there is more direct evidence that LABs are related to massive galaxy formation; luminous sub-mm sources were detected at SSA22 blob1, 2 and No.18 (Chapman et al. 2001, 2003; Smail et al. 2003), while 2142-4420 B1 may not have been observed in sub-mm wavelength. For another example, extended \lya emission is also detected for a SCUBA source SMM 02399-0136 (Ivison et al. 1998; Vernet \& Cimatti 2001).

 The whole nature of LABs at high redshift is still far from understood. How typical are these 100 kpc-scale LABs? What are the luminosity, size, and surface brightness distributions? How are they related to relatively compact \lya emitters (LAEs), or LBGs, or AGNs? It is clear that we need to investigate a larger and systematic sample of LABs in order to understand their nature and relationship with galaxy formation.

 Here we report the properties of 35 robust candidates of LABs at $z=3.1$ in and around the proto-cluster region in the SSA22 field (S00), including SSA22 blob1 and 2, based on wide-field and very deep narrow-band and broad-band images taken with the 8.2m Subaru telescope. The properties of the relatively compact LAEs in the same field were presented in our previous paper (Hayashino et al. 2004; hereafter H04). While H04 also briefly mentioned about the extended \lya halos around LBGs in the field as well as the sky distribution of faint and low surface brightness 'mini blobs' (MBs) based on their eye-ball selection, we here present the results of more objective detection and analysis of these LABs in order to discuss their nature in detail. We use AB magnitudes and adopt a set of cosmological parameters, $\Omega_{\rm M} = 0.3$, $\Omega_{\Lambda} = 0.7$ and $H_0 = 70$ km s$^{-1}$ Mpc$^{-1}$ in this paper. In this cosmology, the universe at $z=3.1$ is 2.0 Gyr old, or 15$\%$ of the present age, and $1''.0$ corresponds to 7.6 kpc of physical length at this redshift.

\section{OBSERVATIONS}

 We obtained wide-field and deep narrow-band ($NB497$) and broad-band  ($B$,$V$, and $R$) images centered at ($\alpha$,$\delta$) = ($22^{\rm h}17^{\rm m}.6$, +00$^{\rm o}17'$) (J2000.0) on 2002 September 8 and 9 (UT) with the 8.2 m Subaru Telescope equipped with the prime-focus camera, Suprime-Cam (Miyazaki et al. 2002). The camera has ten MIT/LL 2048 $\times$ 4096 CCDs arranged in $5 \times 2$ pattern, with the pixel scale of $0''.20$ and the field of view of $34' \times 27'$. Our custom narrow-band filter, $NB497$, has the central wavelength (CW) of 4977\AA\ and FWHM of 77\AA\ to detect \lya emission line at $z = 3.06 - 3.13 $. The spatial variation of the CW and the FWHM of the $NB497$ filter are less than 18\AA\ ($\Delta z=0.015$) and 5\AA\ ($\Delta z=0.004$), respectively. The profiles of the filters are shown in Figure 1. Total exposure times for each filter are listed in Table 1. Typical individual exposure times were 20 minutes for the narrow-band and 6 minutes for the broad-bands with dither motions of more than $30''$ between successive exposures. For $R$-band, an additional 1.3 hours exposure of the same field taken from archive (2000 August and 2001 October, PI: E. M. Hu) was co-added together. 

 The raw data are the same ones as presented in H04 and were reduced with IRAF and a custom software developed for Suprime-Cam data reduction (Yagi et al. 2002; Ouchi et al. 2003) in the similar manner as presented in H04. We made flat fielding using the median sky image and background sky subtraction adopting the mesh size parameter (in Yagi et al's procedure) of $30''$ before combining the images. The final images are slightly different from those used in H04 as we reduced the data by adopting the relatively large background mesh size to avoid the suppression of extended diffuse emission.  All the stacked images were calibrated using spectrophotometric standard stars (Massey et al. 1988; Oke 1990) and Landolt standard stars (Landolt 1992). The magnitudes were corrected for Galactic extinction of $E(B-V)=0.08$, as adopted in S00. 

 The combined images were aligned and smoothed with Gaussian kernels to match their seeing sizes. The average stellar profile of the final images has FWHM of 1.$^{\prime\prime}$0. We constructed a $BV$ image [$BV \approx (2B+V)/3$] for the continuum at the same effective wavelength as the narrow-band filter and an emission line $NB_{\rm corr}$ image by subtracting the $BV$ image from the $NB497$ image. The limiting magnitudes ($1 \sigma$) per square arcsecond are 28.8($NB497$), 29.0($B_{\rm AB}$), 29.0($V_{\rm AB}$), 29.2($R_{\rm AB}$), 29.1($BV$), and 28.8($NB_{\rm corr}$). To evaluate these limiting magnitudes, we fitted a Gaussian function to the distributions of the sky counts that are obtained with apertures of $1^{\prime\prime}.1$ diameter put at random positions in each image and obtained the 1$\sigma$ fluctuation values.  We used only negative parts of the distributions in order to avoid the contamination by the objects except for the case of the $NB_{\rm corr}$ image.

 The total size of the field analyzed here is $31'.1 \times 22'.9$. We masked out the edge region and the halos of the bright stars. The resultant total effective volume probed at $z = 3.1$ by the narrow-band imaging is $1.3 \times 10^5$ Mpc$^3$. 

\section{SELECTION OF \lya BLOBS}

 We obtained a sample of robust candidates of LABs as follows. Object detection and photometry were performed using SExtractor version 2.2.2 (Bertin \& Arnouts 1996). The object detection was made on the $NB_{\rm corr}$ image smoothed with a Gaussian kernel with FWHM of 1$^{\prime\prime}$ and we adopted the criterion, 20 contiguous pixels above the threshold of 6.3 counts per pixel which corresponds to 28.0 mag arcsec$^{-2}$ ($2.2 \times 10^{-18}$ ergs s$^{-1}$ cm$^{-2}$ arcsec$^{-2}$) or to the 2$\sigma$ per square arcsecond of the background fluctuation of the pre-smoothed $NB_{\rm corr}$ image. We adopted the deblending parameter of 0.05 and the background mesh size of $30'' \times 30''$. The magnitudes and colors are measured for each object in the isophotal apertures defined in the process of source detection. In Figure 2, we plot the $BV-NB497$ color versus $NB497$ magnitude diagram for the $NB_{\rm corr}$-selected sources with $NB497\le25$ mag. The solid line shows the limiting color for emission line objects, $BV-NB497=0.7$, which corresponds to an observed equivalent width of 80\AA. In Figure 3, we plot the isophotal area and the $NB_{\rm corr}$ magnitude of the emission line objects with $BV-NB497\ge0.7$. We selected 35 objects with the isophotal area larger than 16 arcsec$^2$ (the solid line), which corresponds to a spatial extent of 30 kpc at $z=3.1$, as the robust candidates of LABs in this paper. We note that the threshold is well above the values for point sources (the dashed line). These 35 candidates of LABs are indicated by the filled squares in Figure 2 and 3.

 We checked the reliability of these 35 LABs on the $NB_{\rm corr}$ image in three different manners. First, we evaluated their significance above the background noise level on the $NB_{\rm corr}$ image assuming that the noise variation does not change significantly with the shape of photometric apertures. We measured $1\sigma$ fluctuation of the sky counts in circular apertures with different diameters of $1''-20''$. We found that all these LABs are significant at more than $8\sigma$ level (the dotted line in Figure 3). Second, we carried out the same detection and selection procedures on the reversed images constructed by multiplying the original images by $-1$. We did not take any false object in this procedure. Finally, we divided the individual frames to the two groups and stacked each to construct the two independent images with half exposure time. We then measured their flux with the same apertures we defined on the total-exposure image. The flux measured on either of the two half-exposure frames are consistent with those on the total one within photometric errors, which supports that the features are not dominated by spurious in some peculiar frames. 

 We can not completely rule out the possibility that our sample is contaminated by [\ion{O}{2}] $\lambda$3727 at $z=0.33$. However, the survey volumes for [\ion{O}{2}] line is 22 times smaller than \lya and there are few known [\ion{O}{2}] emitters with equivalent width larger than 60 \AA\  (e.g., Hogg et al. 1998, Jansen et al. 2000).

\section{RESULTS}

 In Table 2, we summarize the properties of the 35 confident candidates of LABs; we denote them as LAB1, LAB2, ..., and LAB35 in order of the isophotal area. Fourteen of these 35 objects are located in the SSA22a field studied in S00. Of the 14 sources, LAB1 and 2 are the same objects as the blob1 and 2 in S00 and 8 LABs are associated with the known LBGs at $z=3.1$ (Steidel et al. 2003, hereafter S03). These 8 LABs are also indicated by the large open squares in Figure 2, 3, 5, 6, 7, 9, and 10. As presented in H04, we detected 283 relatively compact LAEs and 42 rather diffuse MBs in almost the same field (see H04 for more details). The 283 LAEs were selected with the criteria (i) $NB497<25.8$, (ii) $BV-NB497>1.2$, (iii-a) $B_{\rm AB}-V_{\rm c,AB}>0.2$ for the objects with $V_{\rm c,AB}<26.9$, and (iii-b) $BV-NB497>1.5$ for the objects with $V_{\rm c,AB}>26.9$ using $2''$ aperture photometry. Here, $V_{\rm c,AB}$ represents the emission line free $V$-band magnitude. The 42 MBs are eye-ball-selected objects, which have flux larger than $4\sigma$ in ring apertures with either $2''-3''$ or $2''-4''$. Of these 35 LABs, 17 LABs with bright knots of emission line were included in the LAEs sample and 17 LABs with diffuse halos are also included in the MBs sample in H04. 

 These 35 LABs seem to have a continuous distribution of the isophotal area and emission line flux in Figure 3. LAB1 and 2 are the two most luminous and the largest emitters in our sample; they have isophotal area of 222 and 152 arcsec$^2$, respectively, and total emission line flux brighter than $1 \times 10^{-15}$ ergs s$^{-1}$ cm$^{-2}$. We discovered another notably large emitter, LAB3, which has an area of 78 arcsec$^2$ and flux of $7 \times 10^{-16}$ ergs s$^{-1}$ cm$^{-2}$, both about factor of $\sim 2 - 3$ smaller than LAB1 and 2, while the average surface brightness is a little brighter. The remaining 32 LABs have isophotal area of $16-57$ arcsec$^2$ and flux of $0.7-4.6 \times 10^{-16}$ ergs s$^{-1}$ cm$^{-2}$.

 The morphology and surface brightness profiles of these 35 LABs show a large variety and some of them have interesting and complex structures. Figure 4 shows the $U_n$, $NB497$, $NB_{\rm corr}$, $BV$ and $R$-band images of these LABs. The $U_n$-band images of the SSA22a, b fields (the SSA22a field are shown by dashed line in Figure 9 and the SSA22b field is centered $9'$ south of that position) are taken from the FTP site listed in S03. We show just isophotal areas for the objects outside the SSA22a,b fields. Size of each panel is $25''$ on a side. We adopted the same intensity scaling for all the panels so that their surface brightness in any pass band are compared directly. The yellow lines show the isophotal apertures of the emission line nebulae. The cyan crosses in $R$-band images show the \lya peak positions. The green lines and magenta lines in $R$-band images show the isophotal apertures of the associated LBGs at $z=3.1$, and the nearest continuum sources to the \lya peak, respectively, with the threshold of 28.0 mag arcsec$^{-2}$.  LAB1 and 2 appear to have several `bubble like' structures, which we will examine more in detail in Figure 8 below. LAB5 and 6 show elongated, somewhat conical structures. LAB8 (SSA22a-C15) is located at $15''$ north of LAB1 and is likely to be a part of it. LAB9 and 26 are also closely located with each other with the separation of $\sim 10''$ and there is another compact LAE between them. LAB27 is very clumpy and shows the four knots of emission line within $15''$ separation. LAB18, 20, 25, 26, and 32 appear very diffuse. 

 Note that, while our continuum subtraction works sufficiently, there still remain patches of over or under subtraction at the positions of the bright continuum sources in the panels of $NB_{\rm corr}$ image in Figure 4 (e.g. LAB4, 10, 13, 30, and 33). We examined the effects of these patches in determining the isophotal apertures of these LABs and confirmed that the apertures are well determined avoiding these patches. LAB20 have an over-subtracted patch at the position of the LBG but this can be also caused by \lya absorption in the galaxy. Loss of the line flux by these patches are as small as a few $\%$ of the total in any case. Besides the effects of continuum subtraction, we also note that LAB7 (SSA22a-C6 and M14) is lied close to the halo of a bright star and LAB17 is located near the edge of the image where noise level is relatively high.

 In Figure 5, we plot the isophotal area versus the average surface brightness on $NB_{\rm corr}$ image of these 35 LABs. The solid and dashed lines are the same ones as in Figure 3, but the dotted line shows the average surface brightness 27.6 mag arcsec$^{-2}$, which is a resultant practical detection limit. Although the original detection threshold for 20 pixels (0.8 arcsec$^2$) is 28.0 mag arcsec$^{-2}$, in practice the extended objects with average surface brightness as low as this limit are easily divided into some smaller pieces by noise in the detection procedure. As can be seen in Figure 5, the distribution of the average surface brightness is widespread from relatively compact high surface brightness objects to very diffuse low surface brightness ones. However, there is no source with average surface brightness between 27.0 and 27.6 mag arcsec$^{-2}$ and isophotal area larger than 60 arcsec$^2$. If there are sources in this 'void' of LABs, they should be detected with S/N $> 8$ as shown in Figure 3. It is worth confirming that there is no such largely extended LAB with lower surface brightness than LAB1 and 2 in order to recognize rareness of such gigantic LABs. We checked this by a simple test; the isophotal areas of LAB1 and 2 are still larger than 100 arcsec$^2$, applying 0.7 mag brighter threshold in our source detection. This suggests that if there are LABs with lower surface brightness but with the surface brightness profiles similar to LAB1 and 2, we can detect them as objects with large isophotal area. We also examined how the mesh size in our sky subtraction affects the detection of large and diffuse objects. With our adopted mesh size of 900 arcsec$^2$, flux from the objects up to about 450 arcsec$^2$ is not significantly ($<50\%$) affected by the procedure even for a case of constant surface brightness. In this sense, we also confirmed that LAB1 is the largest Ly$\alpha$ object in this field. 

 In Figure 6, we plot the isophotal area versus the $R$ magnitudes of the continuum sources to see the relation between spatial extents of \lya nebulae and continuum flux. If we select all the continuum sources inside the isophotal area, the $R$ magnitudes must be overestimated. We detected about 43,000 objects down to $R=25.5$ mag in our field of view of $31' \times 23'$. The surface density of the objects is 0.017 arcsec$^{-2}$. The possibility that an unrelated object is inside the halo of $4'' \times 4''$ by chance is as large as 0.3 for each LAB. Therefore, we selected the continuum sources located nearest to the peak of the \lya emission in $R$-band image (i.e. the sources inside the magenta lines in Figure 4) as the candidate counterparts. For the eight objects which are associated with the known LBGs at $z=3.1$, we selected the LBGs as the associated continuum sources (i.e. the sources inside the green lines in Figure 4). The distribution of the $R$ magnitudes of these associates is widespread, especially for smaller LABs and we can not find any correlation in Figure 6. 

 In Figure 7, we plot the isophotal area versus the difference of the position angles (PA) of emission line nebulae and continuum sources. We used the same continuum sources as in Figure 6. The isophotal axes of the continuum sources and the emission line nebulae tend to align with the radio sources for powerful radio galaxies at high redshift (the ``alignment effect'', e.g., McCarthy et al. 1987). In this case, the differences of PA are around 0. If we see bipolar flows from galaxy disks, the differences may be around 90. However, there seems no correlation in Figure 7. 

 In Figure 8, we show the $NB497$ and the $NB_{\rm corr}$ images of the three most luminous and the largest objects in our sample with contour levels of emission line surface brightness to see their internal structures. The contour lines are at $3\sigma$, $6\sigma$ and $9\sigma$ of the sky background fluctuation per square arcsecond (27.6, 26.8, and 26.1 mag arcsec$^{-2}$). Very complicated structures are seen in LAB1 and 2 while rather smooth halo-like structure is seen in LAB3. The position of the CO emission line and sub-mm continuum source in LAB1 is $\Delta$RA=$13''$, $\Delta$Dec=$13''$ in Figure 8 (Chapman et al. 2003). We clearly confirmed the cavity of \lya emission at the position of the CO source (Bower et al. 2004). We can recognize 'bubble' or 'shell'-like round-shape structures in the $NB497$ or $NB_{\rm corr}$ images of LAB1 and 2 (blob1 and 2 in S00); the centers and radii (r) of these bubbles we note here ($\Delta$RA, $\Delta$Dec, r) are (7.2, 17.2, 3.0), (12.4, 7.5, 2.2), (11.6, 14.3, 1.4) and (15.1, 11.1, 1.4) [arcsec] for LAB1 and (10.5, 8.2, 2.5) and (15.0, 15.1, 1.8) for LAB2 in Figure 8. We will speculate about origin of these complicated structures in the next section. 
 
 In Figure 9, we plot the sky distribution of these 35 LABs (Figure 9a) and the 283 relatively compact LAEs in H04 (Figure 9b). The solid lines in both figures show the high density region (HDR) where the local surface number density of the LAEs is larger than the mean value of the entire field. The dashed line shows the field of view of S00 (the SSA22a field). The overall sky distribution of these 35 LABs is very similar to the HDR of the LAEs, which implies that these LABs belong to the same structure of the LAEs. Note that the difference of the limiting magnitudes inside and outside the HDR is only 0.02.

 We also checked whether these candidates are associated with known radio sources or not since such \lya emission line nebulae are often seen around high redshift radio galaxies. None of them is associated with a radio source brighter than the total flux of 1 mJy at 1.4 GHz [the FIRST catalog by White et al. (1997)]. This 1 mJy FIRST detection limit corresponds to $2 \times 10^{32}~{\rm ergs}~{\rm s}^{-1}~{\rm Hz}^{-1}$ at a rest frequency of 6 GHz for objects at $z=3.1$, which is about 2 orders of magnitude fainter than powerful radio galaxies. We also checked the association of X-ray sources and did not find any counterparts in the ROSAT All-Sky Survey (RASS) source catalog (Voges et al. 1999, 2000). The typical limiting flux in the $0.1-2.4$ keV band of the RASS is $\sim 10^{-13}~{\rm ergs}~{\rm cm}^{-2}~{\rm s}^{-1}$, which corresponds to $\sim 10^{46}~{\rm ergs}~{\rm s}^{-1}$ for objects at $z=3.1$. This limit is comparable to the luminosity of the brightest AGNs in X-ray. Deeper X-ray image by {\it Chandra} or {\it XMM-Newton} will help us to investigate presence or absence of AGNs in these LABs.

\section{DISCUSSIONS}

 We first argue that most of the 35 candidates must be LABs at $z=3.1$. This is supported by the large equivalent widths, larger sampled volume than [OII] emitters at $z=0.33$, association with LBGs ($60\%$ in the SSA22a field), and coincidence of overall spatial distribution with that of LAEs selected at further robust equivalent width criteria (H04). 

 In the following part of this section, we discuss the properties of these 35 LABs and consider what their physical origins are and how they are related to galaxy formation process.  

\subsection{\lya blobs and galaxy formation process}

 As discussed in $\S$1, there are at least three possibilities for the origins of these LABs, namely, (i) photo-ionization, (ii) cooling radiation from gravitationally gas in collapsed halos, and (iii) shock heating by starburst driven galactic superwind. Scattering by surrounding neutral gas affects their appearance for all these cases. The case of cooling radiation and superwind represents rather early and late stage of intensive star formation in proto galaxy while photo-ionization by internal sources does on-going events. Sources of photo-ionization may be massive stars or AGNs inside the galaxies, or the background diffuse UV emission. It is interesting to see the objects in the same large-scale structure and may be at different phases of galaxy formation. 

\subsubsection{Photo-ionization}

 Probably, the simplest idea is that we see extended star-forming regions in proto galaxies or diffuse hydrogen gas surrounding or bound to the proto galaxies which is ionized by photons that escaped from galactic star-forming regions. Note that emission line nebulae in our sample seem to be more extended than the continuum sources (see Figure 4) and some of them indeed resemble extended diffuse gas structures of proto-galaxies at high redshift studied in numerical simulation (e.g., Abadi et al. 2003). 

 In order to test this picture, namely, photo-ionization by galactic star-forming regions, we compare the Ly$\alpha$ luminosities with the UV luminosities of the associated objects to see for how many sources the observed UV luminosities are apparently sufficient to produce the \lya emission by photo-ionization. 
 
 In Figure 10, we plot $R$ versus $NB_{\rm corr}$ diagram for these 35 LABs. The filled squares with $1\sigma$ error bars show magnitudes calculated by the same isophotal apertures as used in $\S3$ (Figure 10a). We translated the $NB_{\rm corr}$ and $R$ magnitude into the equivalent star formation rates (SFR) at $z=3.1$, assuming the same Salpeter initial mass function (1955) with mass limits 0.1 and 100 $M_{\odot}$, solar metallicity, no extinction, and case B recombination in the low density limit ($N_{e} \ll 1.5 \times 10^4$ cm$^{-3}$). We used the relationships, $L({\rm Ly}\alpha) = 1.0 \times 10^{42} \times [ SFR / M_{\odot} {\rm yr}^{-1}] ~{\rm ergs} ~{\rm s}^{-1}$ for \lya luminosity (Osterbrock 1989; Kennicutt 1998) and $L_{\nu}({\rm UV}) = 7 \times 10^{27} \times [ SFR / M_{\odot} {\rm yr}^{-1} ] ~{\rm ergs}~ {\rm s}^{-1} ~{\rm Hz}^{-1}$ for UV luminosity (Kennicutt 1998). Of course, the large isophotal apertures suffer from contamination by the foreground and background continuum sources in the $R$-band image and the contribution of UV luminosity in Figure 10a must be overestimated as mentioned in $\S$4. Therefore, to evaluate more reasonable contribution of the UV continuum sources, we also plotted the same diagram in Figure 10b, but selecting the same continuum sources as in Figure 6, namely, associated LBGs or the nearest sources to the \lya peak. In Figure 10a and 10b, six and fourteen of these 35 LABs have $SFR$(Ly$\alpha$)$>SFR$(UV), and thus are apparently not associated with stellar UV continuum sources that are bright enough to produce photo-ionized \lya emission. Namely, for about one third of these 35 LABs, simple photo-ionization by massive stars is not sufficient to explain the \lya luminosities. Note that we neglect a dilution effect of UV continuum flux for this comparison. This means we consider a simple limiting case in which the LAB completely surrounds the continuum source and is optically thick in the Lyman continuum, thereby absorbing all the available ionizing photons. 

 We may consider at least four possibilities to explain these Ly$\alpha$-excess objects which have $SFR$(Ly$\alpha$)$>SFR$(UV) in Figure 10b. The first one is that the UV spectra of these objects are dominated by a stellar population with more massive, metal-poor, and young stars. These Ly$\alpha$-excess objects have the rest \lya equivalent widths, $EW_{\rm rest}$ larger than 100\AA. In order to explain such large \lya equivalent widths, we need to consider more extreme stellar population. Charlot \& Fall (1993) showed that a younger starburst age, or a flatter IMF with high mass cut off are needed to explain a large equivalent width, $EW_{\rm rest}$\simgt 100\AA. For example, the Large Area Lyman Alpha survey found $\approx$ 150 LAEs at $z=4.5$ and 60$\%$ of them have the $EW_{\rm rest}$ larger than 240\AA~ (Malhotra \& Rhoads 2002). They claimed that a stellar IMF with an extreme slope of $\alpha=0.5$ or zero-metallicity stars required to produce such large \lya equivalent widths.

 The second one is that the gas is photo-ionized by AGNs inside forming galaxies (e.g., Haiman \& Rees 2001).  Of the 14 Ly$\alpha$-excess sources, three (LAB23, 24, and 28) have point-like nearest continuum sources which have FWHM smaller than $1''.1$ on the $R$-band image. UV luminous object, LAB21 has also a point continuum source. Comparison with deep X-ray image or optical spectroscopy to detect strong C{\small IV} emission is needed to further constrain the association with AGNs. However, since the nearest continuum sources are not point-like but resolved galaxies in many other cases, we do not think the AGN population dominates our sample.

 The third one is that ionizing UV sources (AGNs or star formation) may be hidden from our line of sight. There are cases that the peaks of \lya emission are not coincident with the continuum sources (LAB9, 13, and 25 for \lya excess objects, and LAB6, 16, 18, and 26 for UV luminous objects). Five of these seven continuum sources have redder $V_{\rm c}-R$ colors ($0.3-0.4$ magnitudes) than those of other sources which are close to the peak of \lya emission ($r\simlt 1''$). If this is due to misidentification of associated continuum sources, {\it true} continuum sources may be further faint objects. Indeed, it is known that sub-mm and CO emission is detected in LAB1 but apart from the position of the known LBG (Chapman et al. 2001, 2003). The amount of hidden star formation for these objects is expected to be from several to 100 $M_\odot$ yr$^{-1}$ as seen in Figure 10, assuming no extinction for \lya emission.

 The last one, especially for very diffuse LABs, is photo-ionization of hydrogen gas in outer part of galaxies by the diffuse inter galactic UV background. Some authors evaluated diffuse Ly$\alpha$ emission from systems ionized by background UV radiation. For example, Gould \& Weinberg (1996) evaluated properties of \lya emission from \lya clouds with neutral hydrogen column density of $10^{17}$--$10^{20}$ cm$^{-2}$. They found that the typical surface brightness of the \lya clouds is $10^{-19}$ ergs s$^{-1}$ cm$^{-2}$ arcsec$^{-2}$ at $z=3$ and this value has a weak dependence on column density. Our detection limit, $\sim 3 \times 10^{-18}$ ergs s$^{-1}$ cm$^{-2}$ arcsec$^{-2}$ is still considerably brighter than this value.

\subsubsection{Cooling radiation from gravitationally heated gas}

 Besides photo-ionization by massive stars or AGNs, we should take the contribution of cooling radiation from gravitationally heated gas in collapsed halos into account (Rees, \& Ostriker 1977, S00, Haiman, Spaans, \& Quataert 2000, Fardal et al. 2001). Since this is a direct consequence of atomic gas cooling process during galaxy formation, it is very important to find objects radiating Ly$\alpha$ emission by this mechanism. Typical flux and size, $\sim 10^{-17}$--$10^{-16}$ erg s$^{-1}$ cm$^{-2}$ and $\sim $5--10 arcsec, (e.g., Table 1 in Haiman et al. 2000) roughly match with the objects discussed in this paper.

 According to the simulation by Fardel et al. (2001), objects with large radiative cooling with $L($Ly$\alpha$$_{\rm cool}$$)$ $> 10^{43}$ ergs s$^{-1}$ also have high star-formation activity and the total \lya luminosities of such objects become larger than 10$^{44}$ ergs s$^{-1}$. Below the luminosities, however, cooling radiation sometimes dominates \lya emission.  Given the luminosity range of our sample of these 35 LABs, $6 \times 10^{42}$ to $10^{44}$ erg s$^{-1}$, it is possible that objects dominated by cooling radiation are contained in our sample. 

 Interestingly, there are several objects which are very diffuse and are not associated with bright continuum sources (LAB5, 8, 9, 13, and 25). They are good candidates of the \lya cooling objects. There are also objects with extended faint 'halo' structure although they are associated with relatively bright continuum sources (LAB6, 7, 11, 12, 18, and 26).

\subsubsection{Superwind}

 We then discuss the case of shock heating by starburst driven galactic superwind. One of the direct evidence for superwind activities is expanding bubbles (e.g., Heckman, Armus, \& Miley 1990). The bubble like structures of LAB1 and 2 (blob1 and 2) resemble the expanding bubble observed in Arp220 with H$\alpha$ imaging (e.g., Heckman et al. 1996). It is noted that the possible bubbles (see $\S4$) have the typical radius of $2''$, corresponding to 15 kpc at $z=3.1$, which is similar to the radius of the bubble in Arp220. Interestingly, it is known that LAB1 is more luminous by a factor of 30 than Arp 220 and has very similar rest-frame optical and far-infrared spectral energy distribution (Taniguchi, Shioya, \& Kakazu 2001). Ohyama et al. (2003) showed that a bright knot in the central part of LAB1 ($\Delta$RA=$11''$ and $\Delta$Dec=$11''$, in Figure 8) has three velocity components separated by $\sim \pm$ 3000 km s$^{-1}$ and two component profiles found both at $1''-2''$ northwest and $1''.5-2''.5$ southeast of the knot. These profiles may represent the expanding shocked shells. We considered star formation activities to produce each bubble assuming the same expanding time scale of $3 \times 10^7$ yr and kinetic energy injection rate of $\sim 10^{43}$ ergs s$^{-1}$ as those of Arp220 (i.e., Heckman et al. 1996). We use the fraction of supernova energy converted to kinetic energy of 0.3 (e.g., Mori, Ferrara, \& Madau 2002) and the number of supernovae per $M_{\odot}$ of stars formed of 0.007, assuming a Salpeter IMF (e.g., Bower et al. 2001). Then we have the supernovae rate of $\sim$1 yr$^{-1}$ and $SFR \sim 150 ~M_{\odot} ~{\rm yr}^{-1}$ for each bubble. Since we have identified at least four bubbles in blob1, it has total SFR more than $\sim$ 600 $M_{\odot} ~{\rm yr}^{-1}$. This is the same order with the expected currently on-going $SFR \ge 500 ~h^{-2} ~M_{\odot} ~{\rm yr}^{-1}$ of LAB1 from sub-mm observation (Chapman et al. 2001).

On the other hand, Bower et al. (2004) argued that the \lya emission of LAB1 is driven by the interaction of outflowing material and the inflow of material cooling in the cluster potential from the chaotic velocity structure they observed. They suggested that the emission line halo around NGC1275 in the Perseus cluster may be a good local analogue to LAB1 although the LAB1 is factor of $\sim 100$ more luminous and has large velocity width. They also argued, however, that the \lya emission is unlikely to be explained by cooling flow phenomenon alone from the low ${\rm L}_{\rm X} / {\rm L}_{{\rm Ly}\alpha}$ ratio. We note that a number of superbubbles occurred rather simultaneously may also explain the observed chaotic velocity structure.

Bower et al. (2004) also showed that two LBGs (SSA22a C11 and C15 in S00) have clear velocity shear patterns in their \lya emission line (LAB1 and LAB8 here). They suggested that the shear patterns are bipolar outflows by superwind.

\subsection{Sky distribution of these LABs}

 An interesting result presented in this paper is that these LABs seem to be correlated with dense environment of LAEs in H04. While 72\% of the LAEs (205/283) are located in the HDR, 86\% of these LABs (30/35) are in the HDR and other 3 are just outside the HDR but may be in the same structure (see Figure 9). We evaluated the probability that more than 30 of 35 sources randomly selected from the 283 LAEs sample are in the HDR by a simulation, and found that it is 5.7$\%$. If we consider the case of 33/35 sources, it is only 0.1$\%$. Thus the distribution of these LABs is likely to be more concentrated into the HDR than that of the 283 LAEs. Furthermore, we note that one-third of these LABs are clustered at the SSA22a field where the two giant blobs (LAB1 and 2) exist.

 What does the strong clustering of these LABs mean? The simplest idea may be, therefore, that these extended and probably gas-rich objects trace the region where galaxy formation preferentially occurs at $z \sim 3$. Previous numerical simulations coupled with semi-analytic galaxy-formation models claimed that the galaxy formation occurs in biased manner at high redshift and concentrations of star-forming galaxies observed at $z \sim 3$ evolve into rich clusters at later epoch (Governato et al. 1998, Kauffman et al. 1999). 

 Since the distribution of more massive dark matter halos is more strongly biased to mass at high redshift in these models, it is also possible that the extended emitters trace more massive objects than LAEs. Unfortunately, our sample of these LABs are not large enough to estimate mass from their spatial clustering properties. We may use, however, their spatial extents to constrain their mass in the case that the extended \lya nebulae are bound to the galaxies. If we assume that the dark matter halos of these LABs were collapsed at $z=3.1$ and the extents of \lya nebulae are smaller than the virial radii, we may evaluate the lower-limit of the mass. The virial mass of a dark matter halo collapsed at a redshift $z$ can be written as $M = 4 \pi R^3_{\rm vir} \rho_{\rm crit}(z) \Delta_{c}(z) / 3$. Here $R_{\rm vir}$ is the virial radius, $\rho_{\rm crit}(z)$ is the critical density at that redshift, and $\Delta_{c}(z)$ is the virial density. We use the fitting formula of Bryan \& Norman (1998) for the virial density taken from the solution to the collapse of a spherical top-hat perturbation, $\Delta_c = 18 \pi^2 + 82x -39 x^2 $, where $x\equiv \Omega_{\rm M}(z)-1$. The average isophotal area of these 35 LABs is 37 arcsec$^2$, which is equivalent to an area of circle with a radius of $3''.4$ or 26 kpc in physical scale at $z=3.1$. With this value, the lower limit of the mass is to be $4 \times 10^{10}~M_{\odot}$. This limit is very similar to the lower limit of average dynamical mass of LBGs at $z\sim 2$ estimated from their rotation curves (Erb et al. 2003) although it is rather small compared with a typical mass of LBGs of $\sim 10^{12}$ $M_{\odot}$ estimated from their clustering properties at $z\sim 3$ (Adelbergaer et al. 1998). 

 Another possibility is that the clustering may be due to an environment which is relatively rich in neutral hydrogen. The density of intergalactic Ly$\alpha$ clouds may be high inside the large-scale structure of star-forming galaxies. Indeed, Adelberger et al. (2003) found that the transmission of Ly$\alpha$ forest at $z=3.1$ in the SSA22a field (i.e., dashed line in Figure 9) is very low. In such an environment, the gas clouds bound to or around galaxies would tend to be observed as extended diffuse Ly$\alpha$ nebulae.

\section{CONCLUSION AND FUTURE PROSPECTS}
 
 We have presented the first large and systematic sample of LABs at high redshift. The 35 LABs show a continuous distribution of the isophotal area and the emission line flux, and a variety of the morphology and the surface brightness. At least three different origins (photo ionization, cooling radiation, superwind) can be considered to explain their properties. In order to reveal their true nature, however, we will need further deep spectroscopic observations to investigate their kinematics properties, ionization and excitation status, and metallicity in the future. Deep and wide-field $U$-band imaging of the field will also be useful in determination of the true associated continuum sources. Further wide-field narrow-band imaging in other fields will prove their clustering properties, especially the relative concentration to the more compact and strong LAEs, with higher statistical accuracy. 

\acknowledgments

 We thank the staff of the Subaru Telescope for their assistance with our observations. This work is partially supported by the grants-in-aid for scientific research of the Ministry of Education, Culture, Sports, Science, and Technology (14540234). The Image Reduction and Analysis Facility (IRAF) used in this paper is distributed by National Optical Astronomy Observatories. U.S.A., operated by the Association of Universities for Research in Astronomy, Inc., under contact to the U.S.A. National Science Foundation.

\bigskip
\newpage
\newpage

\begin{deluxetable}{cccc}
\tabletypesize{\small}
\tablecaption{Summary of Observations \label{tbl-1}}
\tablewidth{0pt}
\tablehead{
 \colhead{Filter} & \colhead{Central Wavelength} & \colhead{Exposure} & \colhead{1$\sigma$ (lim)}\\
 \colhead{} & \colhead{(\AA)} & \colhead{(hours)} & \colhead{(mag/arcsec$^2$)}\\
}
\startdata
 $NB497$ & 4977 & 7.2 & 28.8 \\
 $B$ & 4390 & 1.2 & 29.0 \\
 $V$ & 5420 & 1.8 & 29.0 \\
 $R$ & 6460 & 2.9 & 29.2 \\
\\
\hline
 $BV$ & $4977$ & - & 29.1 \\
 $NB_{\rm corr}$ & $4977$ & - & 28.8 \\
\enddata
\end{deluxetable}

\begin{deluxetable}{lccccccccc}
\tablecolumns{10}
\tabletypesize{\scriptsize}
\tablecaption{Properties of the 35 candidate LABs}
\tablewidth{0pt}
\tablehead{
\colhead{} & \colhead{RA\tablenotemark{a}} & \colhead{Dec\tablenotemark{a}} & \colhead{$NB_{\rm corr}$}  & \colhead{$F$(Ly$\alpha$)\tablenotemark{b}} & \colhead{$L$(Ly$\alpha$)\tablenotemark{c}} & \colhead{Area\tablenotemark{d}} & \colhead{$<$SB$>$\tablenotemark{e}} & \colhead{Notes} & \colhead{}\\
\colhead{} & \colhead{(arcmin)} & \colhead{(arcmin)} & \colhead{(mag)} & \colhead{} & \colhead{} & \colhead{(arcsec$^2$)} & \colhead{} & \colhead{S03} & \colhead{H04\tablenotemark{f}} \\
}
\startdata
LAB1 & 18.0 &  7.9 & 21.06 & 1.3(-15) & 1.1(+44) & 222 & 26.9 & SSA22a-C11 (Blob1) & \\ 
LAB2 & 14.8 &  8.7 & 21.34 & 1.0(-15) & 8.5(+43) & 152 & 26.8 & SSA22a-M14 (Blob2) & \\ 
LAB3 &  9.8 & 10.7 & 21.76 & 7.0(-16) & 5.8(+43) & 78 & 26.5 & & LAE \\ 
LAB4 & 18.2 & 17.3 & 22.22 & 4.6(-16) & 3.8(+43) & 57 & 26.6 & & \\ 
LAB5 & 21.6 & 11.9 & 23.06 & 2.1(-16) & 1.7(+43) & 55 & 27.4 & & LAE,MB \\ 
LAB6 & 26.6 & 20.1 & 23.17 & 1.9(-16) & 1.6(+43) & 42 & 27.3 & & LAE,MB \\ 
LAB7 & 14.3 &  6.7 & 23.23 & 1.8(-16) & 1.5(+43) & 40 & 27.2 & SSA22a-C6,M4 & \\ 
LAB8 & 18.0 &  8.1 & 23.11 & 2.0(-16) & 1.7(+43) & 39 & 27.1 & SSA22a-C15 & MB \\ 
LAB9 & 11.8 & 12.6 & 23.40 & 1.5(-16) & 1.3(+43) & 38 & 27.4 & & LAE,MB \\ 
LAB10 &  9.0 & 21.0 & 22.83 & 2.6(-16) & 2.2(+43) & 34 & 26.7 & & LAE,MB \\ 
LAB11 & 19.4 & 12.7 & 23.76 & 1.1(-16) & 9.1(+42) & 30 & 27.5 & (SSA22a-C47)\tablenotemark{g} & MB \\ 
LAB12 & 16.6 & 12.2 & 23.83 & 1.0(-16) & 8.6(+42) & 29 & 27.5 & SSA22a-M28 & \\ 
LAB13 &  7.6 & 12.0 & 23.61 & 1.3(-16) & 1.0(+43) & 28 & 27.3 & & MB \\ 
LAB14 & 15.6 & 11.2 & 23.50 & 1.4(-16) & 1.2(+43) & 27 & 27.1 & & LAE,MB \\ 
LAB15 &  7.5 &  5.6 & 22.81 & 2.7(-16) & 2.2(+43) & 26 & 26.4 & & 2 LAEs \\ 
LAB16 & 18.3 &  6.5 & 23.67 & 1.2(-16) & 9.9(+42) & 25 & 27.2 & & LAE,MB \\ 
LAB17 &  0.6 &  2.6 & 23.29 & 1.7(-16) & 1.4(+43) & 24 & 26.8 & & LAE \\ 
LAB18 & 17.3 &  3.1 & 24.14 & 7.8(-17) & 6.4(+42) & 22 & 27.5 & & MB \\ 
LAB19 & 19.6 & 13.9 & 23.40 & 1.5(-16) & 1.3(+43) & 21 & 26.8 & & LAE \\ 
LAB20 & 15.7 &  8.0 & 24.14 & 7.8(-17) & 6.4(+42) & 21 & 27.5 & SSA22a-C12 & MB \\ 
LAB21 &  5.3 &  7.4 & 23.92 & 9.5(-17) & 7.9(+42) & 20 & 27.2 & & \\ 
LAB22 & 15.8 & 18.7 & 23.97 & 9.1(-17) & 7.6(+42) & 20 & 27.2 & & \\ 
LAB23 &  7.6 & 18.4 & 23.75 & 1.1(-16) & 9.2(+42) & 19 & 27.0 & & LAE,MB \\ 
LAB24 &  9.4 &  9.9 & 23.98 & 9.0(-17) & 7.5(+42) & 19 & 27.2 & & LAE \\ 
LAB25 & 18.9 & 11.0 & 24.23 & 7.2(-17) & 5.9(+42) & 19 & 27.5 & & MB \\ 
LAB26 & 12.0 & 12.7 & 24.20 & 7.4(-17) & 6.1(+42) & 18 & 27.4 & & MB \\ 
LAB27 & 22.7 & 16.6 & 24.10 & 8.1(-17) & 6.7(+42) & 18 & 27.3 & & MB \\ 
LAB28 &  9.8 & 18.0 & 22.83 & 2.6(-16) & 2.2(+43) & 18 & 26.0 & & LAE \\ 
LAB29 & 26.0 & 18.1 & 24.08 & 8.2(-17) & 6.8(+42) & 17 & 27.2 & & MB \\ 
LAB30 & 16.4 &  6.8 & 23.72 & 1.1(-16) & 9.5(+42) & 17 & 26.8 & SSA22a-D3 & \\ 
LAB31 & 14.8 &  6.3 & 23.55 & 1.3(-16) & 1.1(+43) & 17 & 26.6 & SSA22a-C16 & LAE \\ 
LAB32 & 18.5 & 17.1 & 24.26 & 7.0(-17) & 5.8(+42) & 17 & 27.3 & & LAE,MB \\ 
LAB33 &  6.5 &  9.8 & 23.78 & 1.1(-16) & 9.0(+42) & 16 & 26.8 & & \\ 
LAB34 & 24.9 & 19.6 & 23.99 & 9.0(-17) & 7.4(+42) & 16 & 27.0 & & LAE \\ 
LAB35 & 18.3 & 12.5 & 23.78 & 1.1(-16) & 9.0(+42) & 16 & 26.8 & & LAE \\ 
\enddata
\tablenotetext{a}{Using the coordinate system shown in Figure 9.}
\tablenotetext{b}{\lya emission line flux in the unit of ergs s$^{-1}$ cm$^{-2}$.}
\tablenotetext{c}{\lya luminosity at $z=3.1$ in the unit of ergs s$^{-1}$.}
\tablenotetext{d}{Isophotal area determined on the $NB_{\rm corr}$ image.}
\tablenotetext{e}{Average surface brightness on the $NB_{\rm corr}$ image in the unit of mag arcsec$^{-2}$}
\tablenotetext{f}{LAE and MB stand for \lya emitter and mini-blob selected in H04.}
\tablenotetext{g}{No redshift information in S03.}
\end{deluxetable}

\begin{figure}
\figurenum{1}
\epsscale{0.8}
\plotone{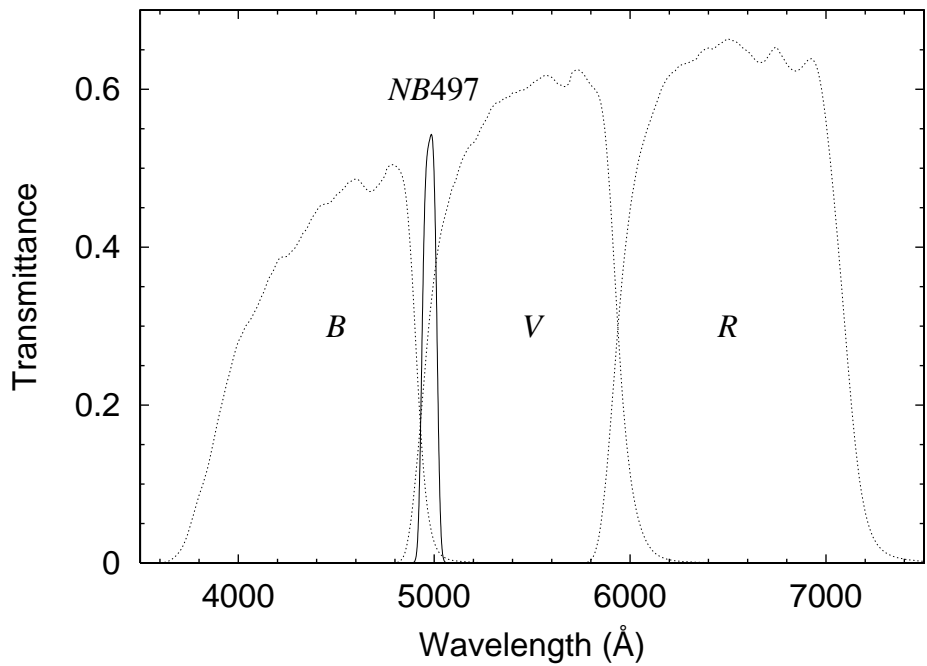}
\caption{ Transmittance of the $NB497$ (solid line), $B$, $V$, and $R$ band (dashed lines) filters for F/1.86 beam of Suprime-Cam. The profiles include the CCD quantum efficiency of Suprime-Cam, the transmittance of prime focus corrector, and the reflectivity of primary mirror of Subaru telescope.}
\end{figure}

\begin{figure}
\epsscale{1}
\figurenum{2}
\plotone{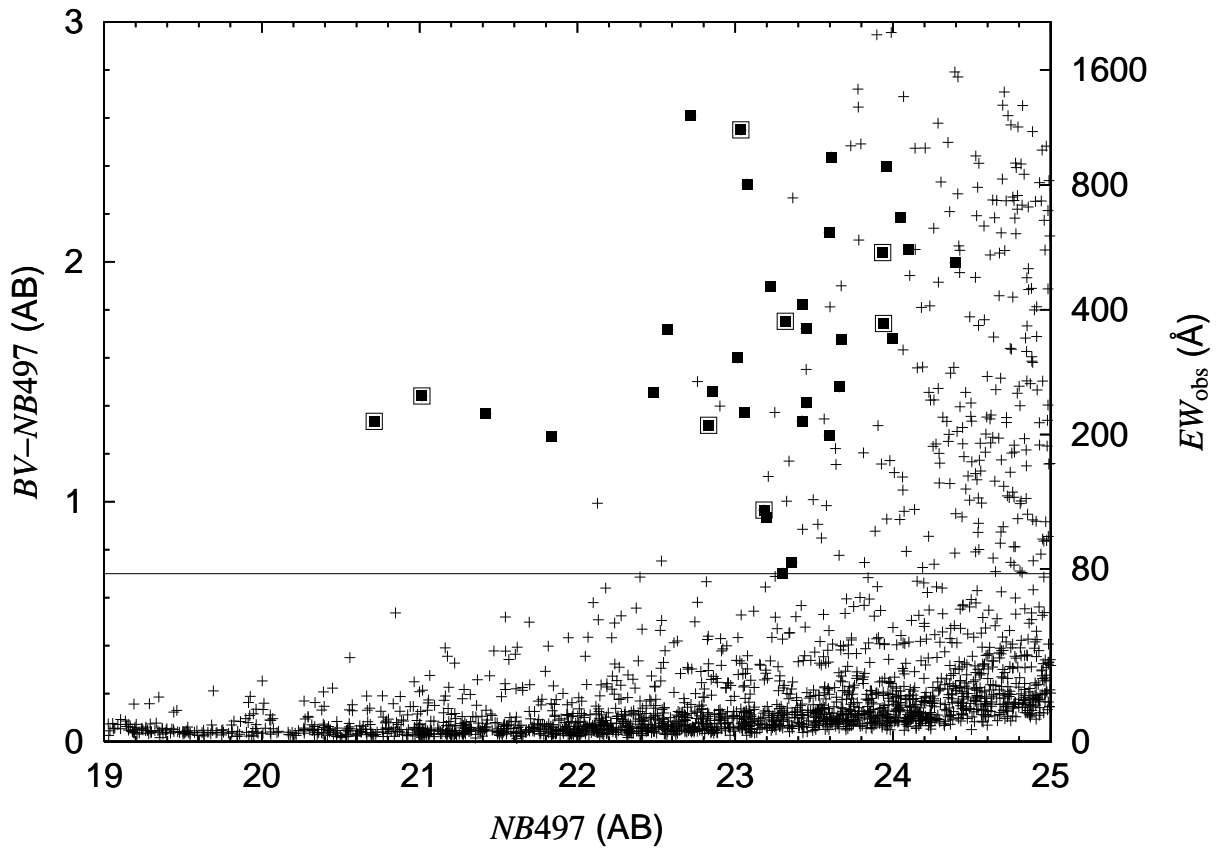}
\caption{ Color-magnitude diagram for $BV-NB497$ color and $NB497$ magnitude. The solid line shows a color of $BV-NB497=0.7$, which corresponds to observed equivalent width of 80\AA\ . The filled squares show candidates of LABs. The large open squares show the candidates which are associated with known LBGs at $z=3.1$ in the SSA22a field (S03).}
\end{figure}

\begin{figure}
\epsscale{0.9}
\figurenum{3}
\plotone{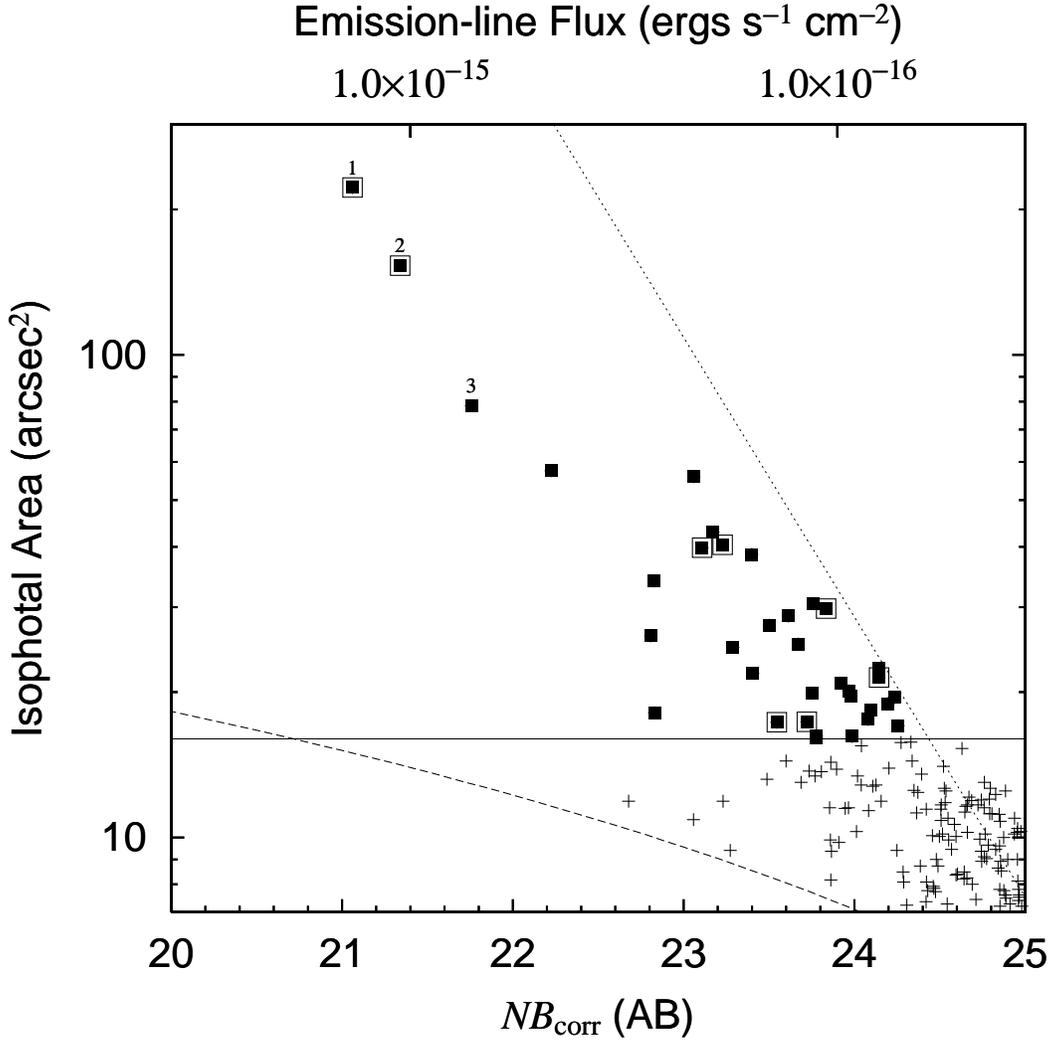}
\caption{ Distribution of isophotal area and magnitude on the $NB_{\rm corr}$ image for candidates of emission line objects brighter than $NB497=25.0$. We selected the objects with the isoarea larger than 16 arcsec$^2$ (solid line) as candidates of LABs (filled squares).  The dashed line shows the expected value for point sources. The dotted line shows 8 $\sigma$ noise level of $NB_{\rm corr}$ magnitude for a given area. The large open squares show the candidates which are associated with known LBGs at $z=3.1$ in the SSA22a field (S03).}
\end{figure}

\begin{figure}
\figurenum{4}
\caption{ $Un$ ( or just isophotal areas ), $NB497$, $NB_{\rm corr}$, $BV$ and $R$ images of the 35 candidates of LABs. Each panel is $25''$ square with the candidate centered. The yellow lines show the isophotal apertures. The cyan crosses, green lines, and magenta lines in $R$-band images show the \lya peak positions, the associated LBGs at $z=3.1$, and the nearest continuum sources to the \lya peaks, respectively.}
\end{figure}

\begin{figure}
\epsscale{0.9}
\figurenum{5}
\plotone{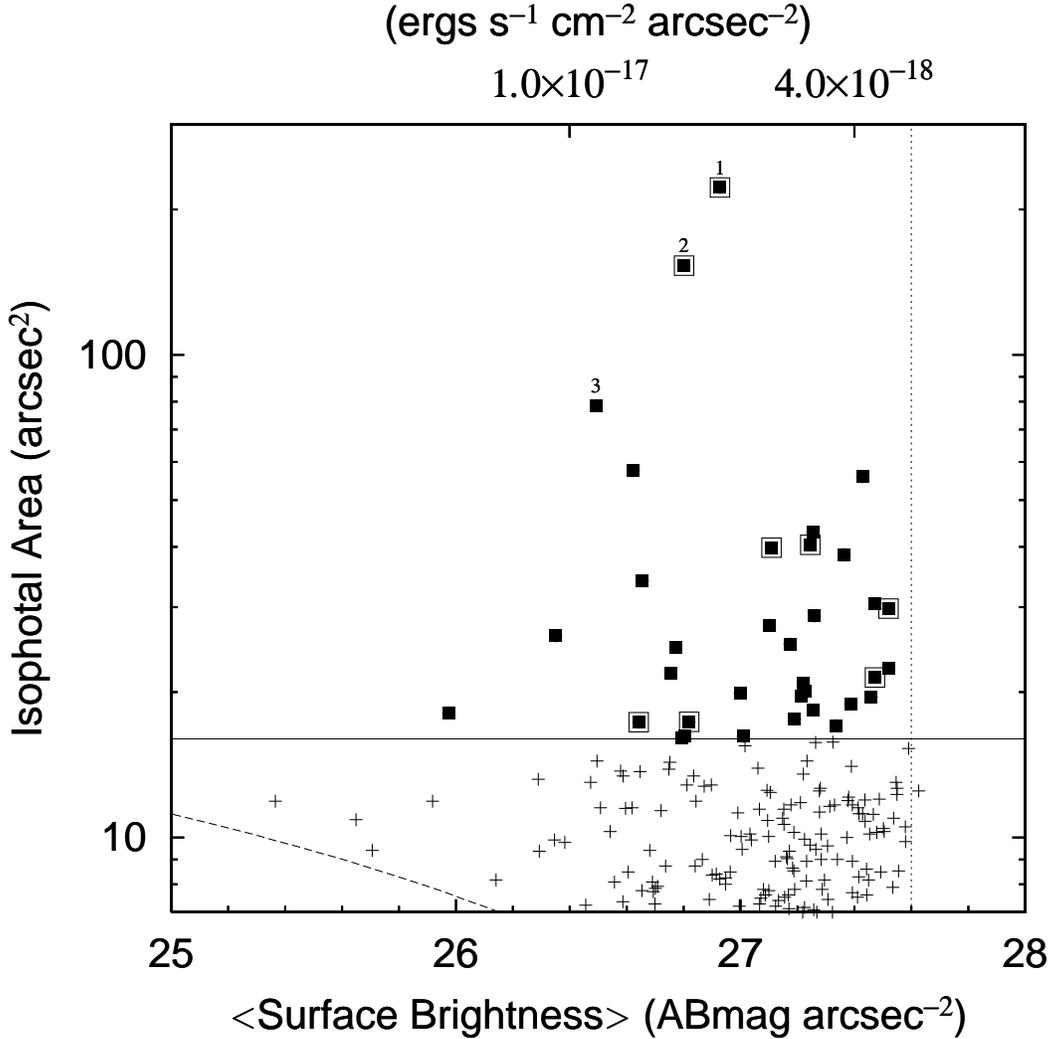}
\caption{ Distribution of isophotal area and average surface brightness on the $NB_{\rm corr}$ image for the 35 candidates of LABs (filled squares). The solid line shows the sample threshold of the isophotal area for LABs (16 arcsec$^2$). The dotted line shows the resultant practical detection limit of average surface brightness $\sim$27.6 mag arcsec$^{-2}$. The dashed line shows the expected distribution for point sources. The large open squares show the candidates which are associated with known LBGs at $z=3.1$ in the SSA22a field (S03).}
\end{figure}

\begin{figure}
\epsscale{0.9}
\figurenum{6}
\plotone{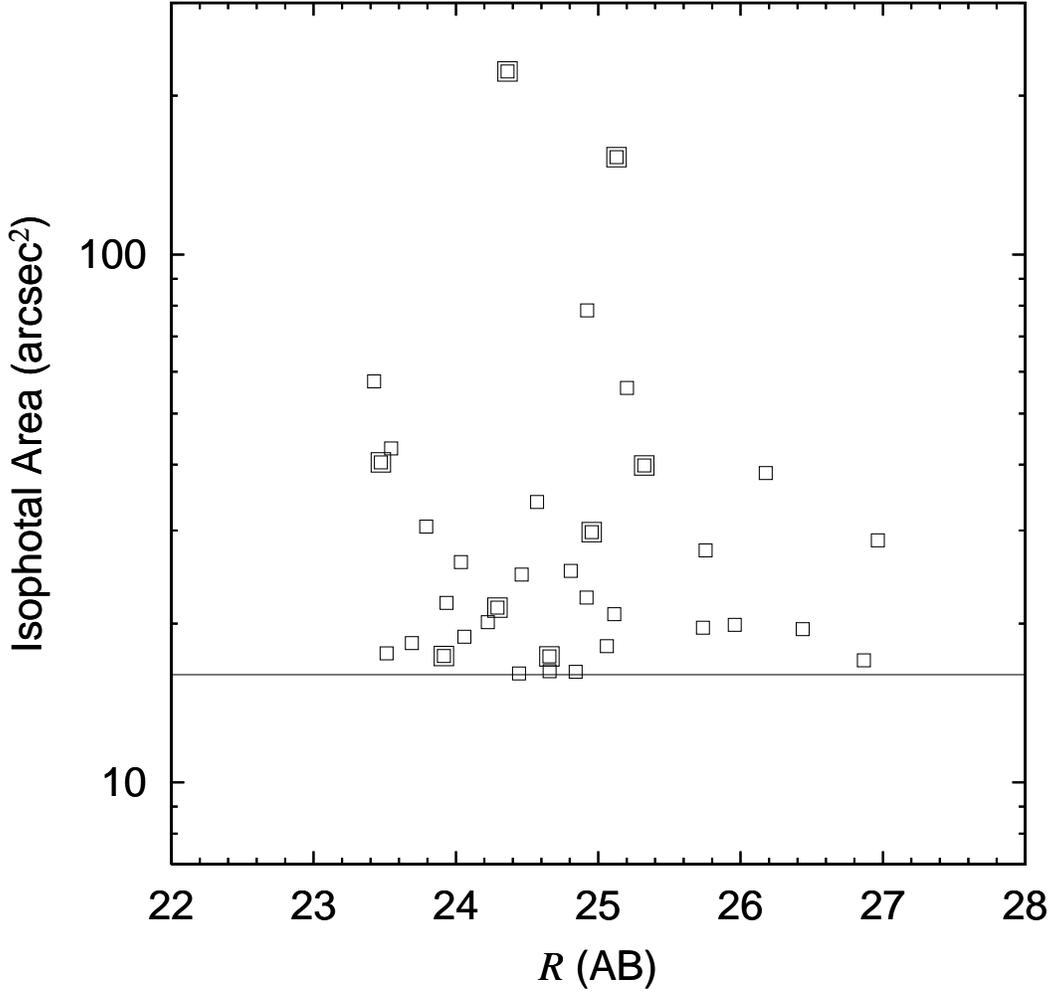}
\caption{Distribution of isophotal area versus $R$ magnitudes of the continuum sources. We selected the continuum sources located nearest to the peak of the \lya emission in $R$-band image (i.e. the sources inside the magenta lines in Figure 4). For the eight objects which are associated with the known LBGs at $z=3.1$, we selected the LBGs as the continuum sources (i.e. the sources inside the green lines in Figure 4). The solid line shows the threshold of the isophotal area for LABs (16 arcsec$^2$). The large open squares show the candidates which are associated with known LBGs at $z=3.1$ in the SSA22a field (S03).}
\end{figure}

\begin{figure}
\epsscale{0.9}
\figurenum{7}
\plotone{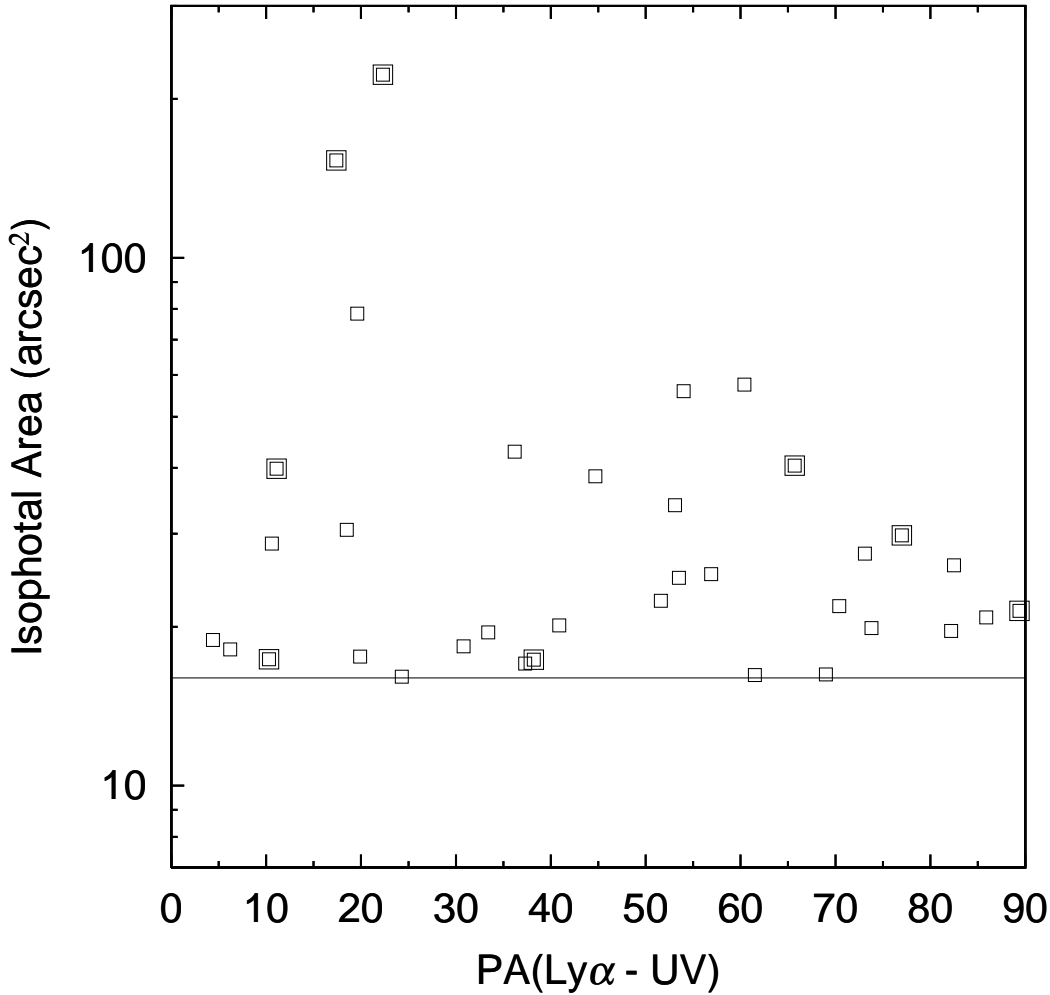}
\caption{Distribution of isophotal area and difference of position angles of the emission line nebulae and the continuum sources. We used the same continuum sources as in Figure 6. The solid line shows the threshold of the isophotal area for LABs (16 arcsec$^2$). The large open squares show the candidates which are associated with known LBGs at $z=3.1$ in the SSA22a field (S03).}
\end{figure}

\begin{figure}
\figurenum{8}
\caption{ $NB$ and $NB_{\rm corr}$ images of the 3 most luminous and largest LABs in our sample. The contour lines are at levels of $3\sigma$, $6\sigma$ and $9\sigma$ per square arcsecond (27.6, 26.8 and 26.1 mag arcsec$^{-2}$).}
\end{figure}

\begin{figure}
\epsscale{0.8}
\figurenum{9}
\plotone{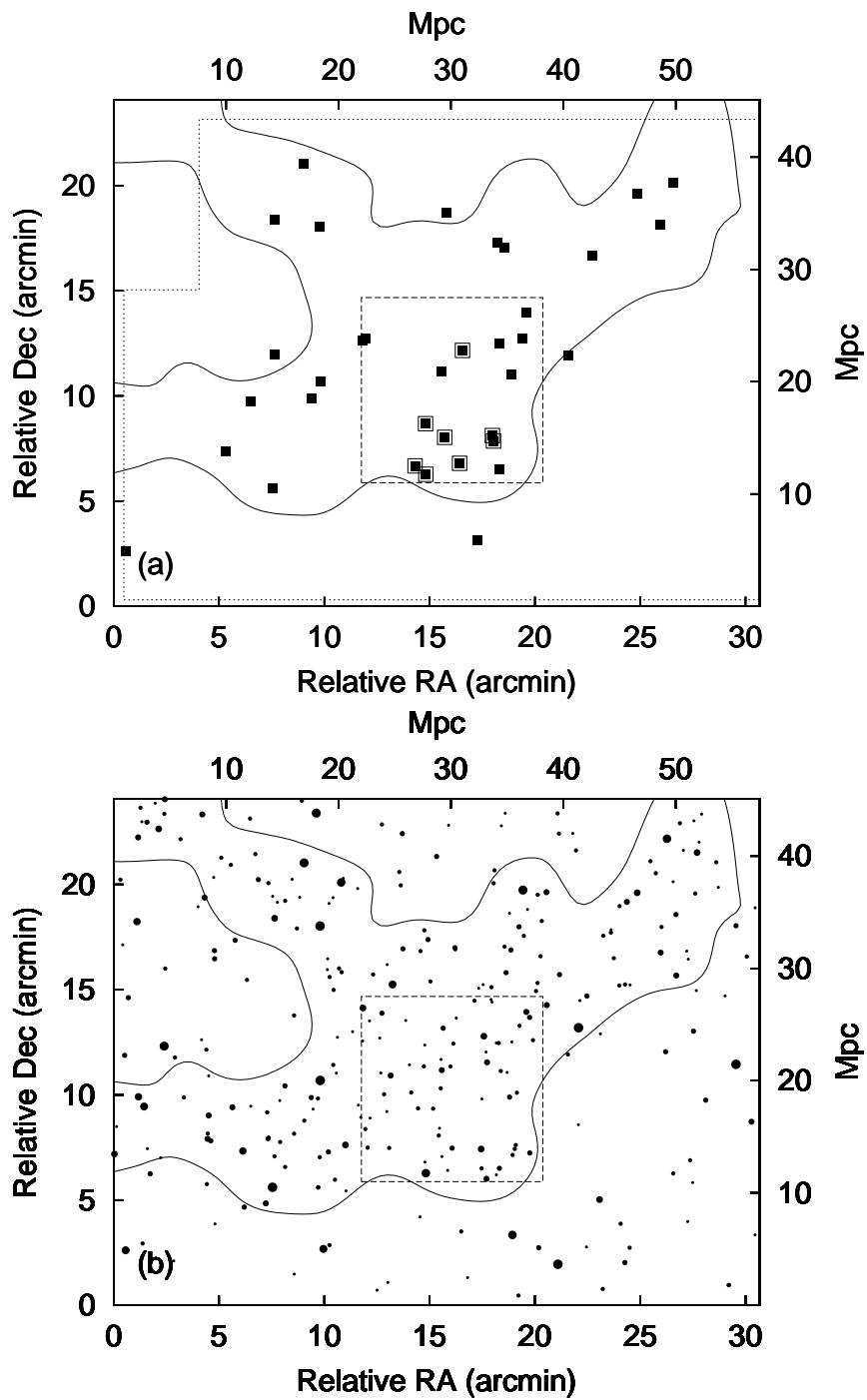}
\caption{ Spatial distribution of the 35 candidates of LABs (top panel) and the relatively compact 283 LAEs presented by Hayashino et al. (2004) (bottom panel). The solid contours show the high density region (HDR) of the 283 LAEs. The dotted line in top panel shows the region which we use in this paper. The dashed line shows the SSA22a field and the large open squares show the candidates which are associated with known LBGs at $z=3.1$ (S03).}
\end{figure}

\begin{figure}
\epsscale{1}
\figurenum{10}
\plotone{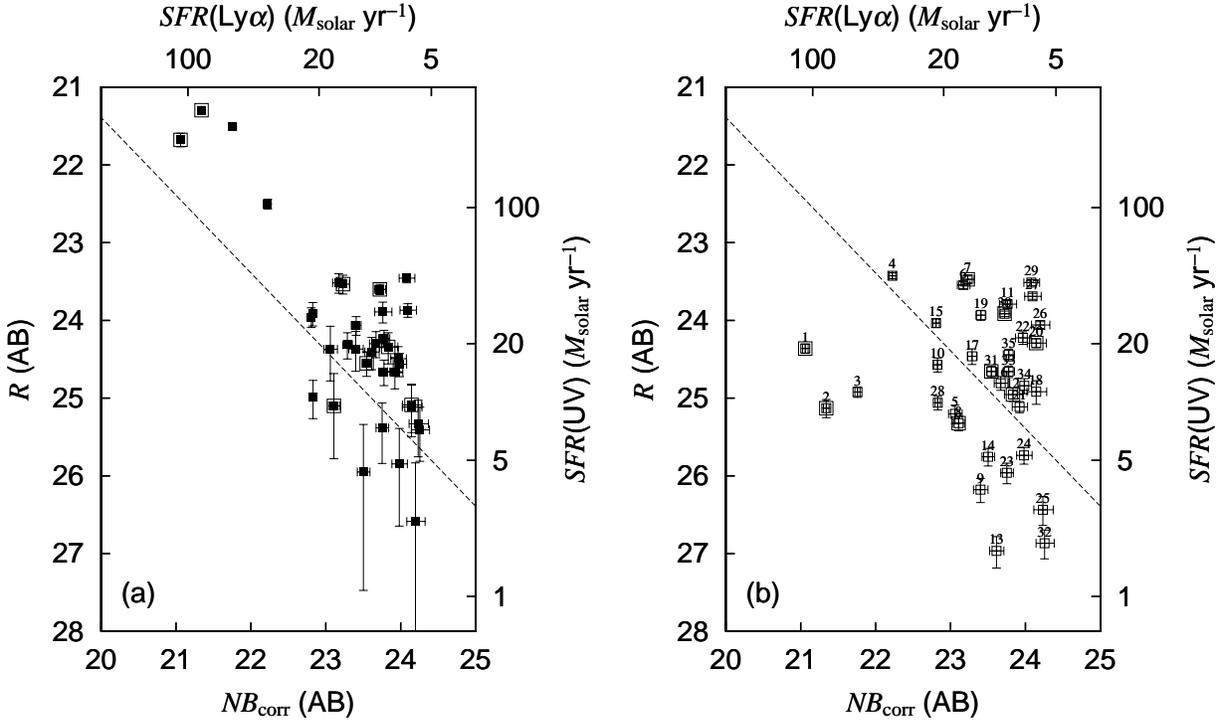}
\caption{ Distributions of $R$ and $NB_{\rm corr}$ magnitudes for the 35 candidates of LABs. In Figure 10a (left panel), the magnitudes are measured on the same isophotal apertures determined with $NB_{\rm corr}$ image. In Figure 10b (right panel), the $NB_{\rm corr}$ magnitudes are the same as left panel, but $R$ magnitudes are those of continuum sources nearest to the \lya peak or the known LBGs at $z=3.1$. The dashed line shows $SFR({\rm Ly}\alpha) = SFR({\rm UV})$. The large open squares show the candidates which are associated with known LBGs at $z=3.1$ in the SSA22a field (S03).}
\end{figure}

\end{document}